# The Great Chicken-and-Egg of Chemistry: Bonding *vs*. Stability Revisited

Chérif F. Matta*[1-4]

[1] *Department of Chemistry and Physics, Mount Saint Vincent University, Halifax, Nova Scotia, Canada B3M 2J6.* [2] *Department of Chemistry, Saint Mary's University, Halifax, Nova Scotia, Canada B3H 3C3.* [3] *Department of Chemistry, Dalhousie University, Halifax, Nova Scotia, Canada B3H 4J3.* [4] *Dép. de chimie, Université Laval, Québec, Québec, Canada G1V 0A6.*

* Correspondence: cherif.matta@msvu.ca

**GRAPHICAL ABSTRACT**

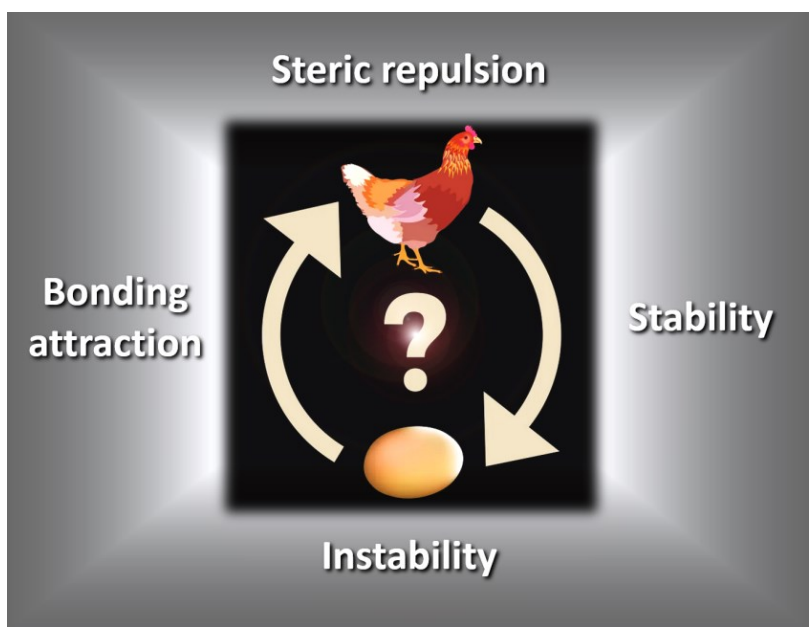

**Graphical Abstract:** The "chicken-and-egg" problem of chemical bonding and stability: bonding attraction and steric repulsion are commonly invoked as causes of stability or instability, yet both are descriptors derived from the underlying quantum state. Treating them as independent causal agents leads to circular or asymmetrical reasoning.

**Abstract**

The chemical bond is a central organizing concept in chemistry, yet it is absent from the molecular Hamiltonian and no "bond operator" exists. Bonding is therefore not a primitive physical entity but a derived descriptor emerging from the quantum state. The logical consequences of this observation are revisited. Statements such as "*bonding stabilizes structure*" when taken literally risk circular reasoning (*petitio principii*), whereby bonding is inferred from a stationary structure and then invoked as its cause. The same caution applies to concepts such as steric repulsion, which is also a derived descriptor. Bonding *accompanies* stable or metastable states and *correlates* with their properties without constituting their cause. Illustrative examples are drawn from QTAIM, non-covalent interaction (NCI) approach, protein structure, and hydrogen–hydrogen bonding. Causation, language, and the autonomy of chemistry are also briefly discussed. The aim is *not at all* to diminish the role of bonding, but to place it at the correct logical level, that is, as a powerful, state-dependent descriptor that organizes, classifies, and predicts chemical behavior without serving as its fundamental cause.

**Keywords:**   Chemical bonding. Stability and reactivity. QTAIM. Noncovalent interactions (NCI). Causation in chemistry. Emergence and reduction.

**Introduction**

A common statement in chemistry is that bonding stabilizes structure. We read that proteins are stabilized by hydrogen bonds, molecular crystals are stabilized by non-covalent interactions, transition states are destabilized by weakened bonds, *etc*. For example, Creighton on page 300 [1] in a section titled "*Rationalizing Stabilities of Folded Conformations*" states that "*The folded structures of proteins indicate that they are likely to be stabilized by a combination of hydrogen bonds, van der Waals interactions, electrostatic interactions, and the hydrophobic interaction.*" On page 142 in Garrett and Grisham [2] underneath the title "*What Noncovalent Interactions Stabilize the Higher Levels of Protein Structures?*" we read that "*… higher levels of structure-secondary, tertiary, and quaternary- are formed and stabilized by weak, noncovalent interactions …*". McMurry [3] asserts that that "*The enol can only predominate when stabilized by conjugation or intra molecular hydrogen bonding*" on page 728, while on page 861 we read that "*Different cellulose molecules then interact to form a large aggregate structure held together by hydrogen bonds*", and on page 893 that "*The structure is stabilized by hydrogen bonds between amide N–H groups and C=O groups four residues away, with an N–H····O distance of 2.8 Å.*" In the International Tables of Crystallography we encounter "*… [hydrogen bonds] play a critical role in the structure and function of proteins and nucleic acids. … Local interactions are very common for main-chain···side-chain hydrogen bonds and may help direct protein folding.*" [4].

    Bonding is one of the most powerful concepts in chemistry. It compresses enormous structural, energetic, and spectroscopic information into a manageable language. It allows chemists to predict structures, explain trends, communicate mechanisms, and design molecules. Meanwhile, chemists speak as if bonds do things: Bonds hold atoms together, hydrogen bonds stabilize proteins, covalent bonds determine molecular skeletons, metallic bonds explain electrical conductivity, noncovalent interactions drive molecular recognition, *etc*. Such statements are useful, but they are also dangerous if taken literally. They encourage the reification of a descriptor into a cause. Because bonding language "works" the bond is effectively turned into a primitive object or a primitive causal agent, and that is the conceptual danger.

    The quantum molecular Hamiltonian contains no bond. It contains masses, charges, kinetic energies, and Coulomb interactions and nothing else. The chemical bond is not an ingredient of the theory underpinning the very chemical bond. The bond is an *emergent* concept inferred after solving, approximating, or analyzing the quantum problem. This point is elementary, but its consequences are often ignored.

    The nonrelativistic Hamiltonian for electrons and nuclei in absence of external electromagnetic fields may be written, in standard notation, as:

$$\hat{H} = -\sum_i \frac{\hbar^2}{2m_e}\nabla_i^2 - \sum_A \frac{\hbar^2}{2M_A}\nabla_A^2 - \sum_{iA}\frac{Z_A e^2}{4\pi\varepsilon_0 r_{iA}} + \sum_{i<j}\frac{e^2}{4\pi\varepsilon_0 r_{ij}} \quad (1)$$
$$+ \sum_{A<B}\frac{Z_A Z_B e^2}{4\pi\varepsilon_0 R_{AB}}.$$

This expression contains no operators such as $\hat{B}_{\text{(non)covalent}}$, $\hat{B}_{\text{hydrogen}}$, $\hat{B}_{\text{bond}}$, $\hat{B}_{\text{metallic}}$, $\hat{B}_{\text{steric repulsion}}$, *etc*. In the presence of external electric or magnetic fields, matter–field coupling terms are added which can result in a change of the quantum state and its associated bonding descriptors, but they still do not introduce a primitive bond operator into the Hamiltonian.

The Hamiltonian codes for kinetic energy and electromagnetic interactions. Meanwhile, Dirac's definition of "observables" requires operators associated with measurable quantities; bonding is not represented by such an operator [5]. Bader's quantum theory of atoms in molecules (QTAIM) gives a real-space description of atoms in molecules and their bonding interactions, but even in QTAIM the bond path (as opposed to the "bond") is a derived emergent topological feature and not a primitive term [6–8].

My point in this essay has several antecedents. For example, Hendry distinguishes structural and energetic conceptions of the chemical bond [9]. Weisberg challenges the structural conception of bonding and emphasized the difficulty of treating the covalent bond as a precisely defined submolecular object [10]. Needham enquires about the source of chemical bonding and warns against simple explanatory language that hides rather than solves the conceptual issue [11]. Esser argues that QTAIM supports an interactive conception of bonding rather than an object-like one [12]. Recently, Seifert has defended the view that the chemical bond is a "real pattern" rather than a simple object [13]. The thesis advanced here is narrower, and that is, that a chemical bond is an emergent higher-level descriptor of a quantum state which is a useful predictor in chemistry, but it is not a primitive cause of stability. Bonding and stability are *not* related by a simple cause and effect chain. Rather, they are two different descriptions of the same underlying quantum state.

**The logical order: Hamiltonian → state → stability → descriptor**

The Born–Oppenheimer approximation introduces structure and the concept of a potential energy hypersurface into chemistry. For fixed nuclear coordinates **R**, one obtains an electronic state $\Psi_\mathbf{R}$ and an energy

$$E(\mathbf{R}) = \langle \Psi_\mathbf{R} | \hat{H}(\mathbf{R}) | \Psi_\mathbf{R} \rangle. \tag{2}$$

An equilibrium structure (a minimum, or a transition state of any order) corresponds to a stationary point on the potential-energy hypersurface:

$$\nabla_\mathbf{R} E(\mathbf{R}_*) = 0, \tag{3}$$

with positive curvature along the stable nuclear-displacement directions, where the star in the symbol $\mathbf{R}_*$ means a special value of **R**, specifically, a stationary-point geometry, and where **R** represents the set of all nuclear coordinates, *i.e.*, $\mathbf{R} = \{\mathbf{R}_1, \mathbf{R}_2, \ldots, \mathbf{R}_N\}$.

Only after this last step do we analyze the resulting state in terms of descriptors such as bonds, bond orders, bond paths, delocalization indices, localized orbitals, NCI surfaces, ELF basins, or energy-decomposition components. A descriptor

$$D = D[\Psi_{\mathbf{R}*}], \tag{4}$$

or

$$D = D[f_{\mathbf{R}*}], \tag{5}$$

is a functional of the state or of a field $f$ obtained from the state, *e.g.*, the electron density, an energy density, a force density etc. It is not the state's primitive cause.

The logical direction is therefore

$$\widehat{H} \to \Psi \to \rho, \Gamma, E(\mathbf{R}) \to \mathbf{R}_* \to D[\Psi] \text{ or } D[f]. \tag{6}$$

The reverse direction

$$D \to \mathbf{R}_* \to E(\mathbf{R}) \to \Psi \to \widehat{H} \tag{7}$$

is not fundamental.

Chemists often reason heuristically in the reverse direction (Relation (7)). That is acceptable as shorthand but not acceptable as an ontology. This is the core of the chicken-and-egg fallacy whereby one first identifies a stable or metastable structure, then one assigns bonding, and then one says that the assigned bonding stabilizes the structure. *Unless the assigned bonding descriptor can be varied independently of the underlying quantum state, the reasoning risks circularity*. The descriptor is inferred from the state and then invoked as the cause of the state. That is the *petitio principii*.

Woodward advocates an interventionist approach where one would need a variable that can be manipulated independently while the relevant background conditions are controlled [14]. A chemical bond is not such a variable. To change a bond is to change the wavefunction, density, geometry, external potential, or environment from which the bond was inferred. The bond is not an independently tunable knob. It is part of the description of the changed state.

This does not invalidate chemical explanation; it simply asks to refine it. Statements such as "hydrogen bonds stabilize the protein" should be read as compressed comparative claims that, in a family of related conformations, patterns classified as hydrogen bonding correlate with lower free energy, altered barriers, or altered structural persistence.

Fig. 1 summarizes the logical sequence underlying the Born–Oppenheimer molecular structure and the assignment of bonding descriptors. For a given nuclear configuration $\mathbf{R}$, the electronic Schrödinger equation $\widehat{H}\Psi = E\Psi$ is solved to obtain the energy $E(\mathbf{R})$. The geometry is then iteratively varied until the stationary-point condition is satisfied (Eq. (3)). The resulting structure $\mathbf{R}_*$ corresponds to a stationary point on the potential energy hypersurface whether a minimum (stable structure) or a saddle point (transition state). Only at this stage are chemical descriptors {D} such as hydrogen bonds, bond paths, or other bonding constructs defined. These descriptors are derived quantities, obtained *after* the determination of the stationary point. The frequently encountered interpretations in which such descriptors are subsequently invoked as the cause of the stability or instability of the structure constitutes a logical inversion of this sequence and amounts to circular reasoning in the first case and asymmetric reasoning in the second.

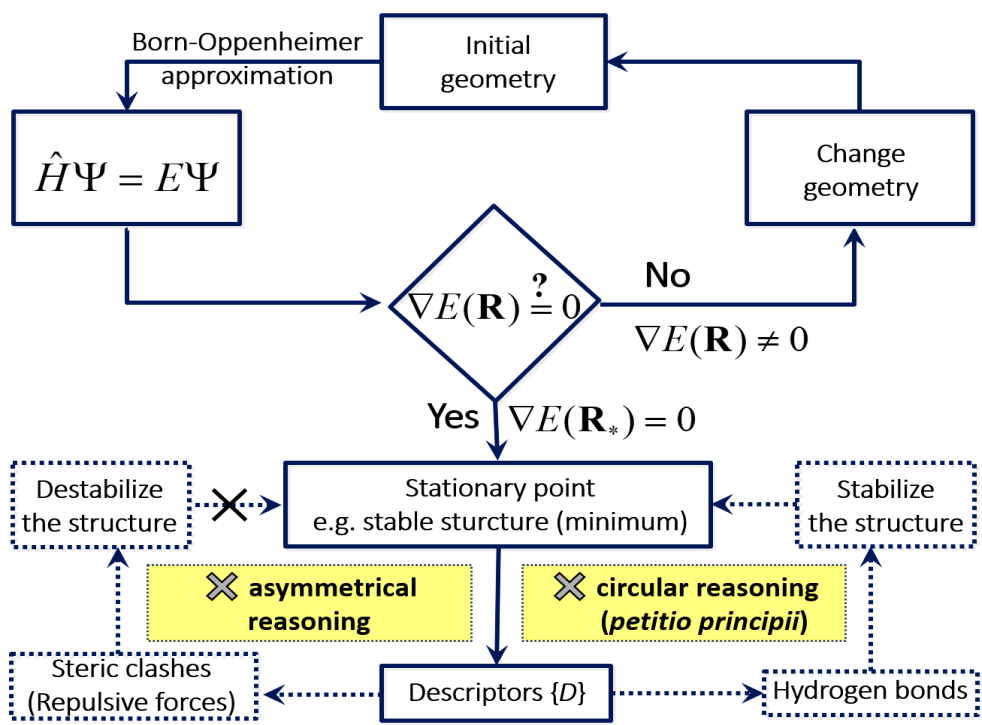

**Fig. 1.** Logical flow to chemical bonding descriptors starting from the of the Born–Oppenheimer approximation and an initial nuclear geometry. Chemical descriptors {*D*} e.g., hydrogen bonds, steric repulsions, bond paths, ELF, EDA, NCI, etc. are assigned to a stationary point structure. The dashed pathway on the right highlights the logical inconsistency of invoking such descriptors as *causes* of the stability of the structure from which they were themselves derived. The other dashed pathway on the bottom left centered on destabilizing descriptors, *e.g.*, steric repulsion, which are similarly treated as causal agents of instability leading to asymmetrical reasoning (also known as "*ad hoc* causal attribution"), *i.e.*, selectively assigning causation after the fact.

**Electron density, QTAIM, and the real-space recovery of chemistry**

Bader's quantum theory of atoms in molecules (QTAIM) distinguises the bond path, a line of maximal electron density in space that can be curved, from a "bond" depicted as a "little straight rod" between atoms. Bader's warns us that "*bond paths are not chemical bonds*" [7]. Bader's point is that *a descriptor should not be reified*, and that a bond path indicates a special interaction between atoms. Reinforcing the interaction concept, the bond path has been shown homeomorphic with a virial path, a line in real space connecting two nuclei along which the potential energy density $V(\mathbf{r})$ (the virial field) is locally maximum [15]. The same bond path has also been shown as a "*privileged exchange channel*" (a one-electron feature capturing a two-electron information) (See Ref. [16]).

This is why Bader insisted on using the verb "*bonding*" instead of the noun "*bond*", and this nuance *does* matter. Esser's discussion of QTAIM as an interactive conception of bonding further elaborates the reasons to avoid a crude object ontology [12]. In 2017, I suggested to use the word "*associated*" when connecting energetic stability and chemical bonding to avoid causality implications [17].

**Virial paths, energy density, and why association is not causation**

An objection may be that if every bond path is mirrored by a virial path [15], and if the virial path is a line of maximally stabilizing potential energy density, does that not show that bonding causes stability? The short answer is: *no*, it does not. What this shows is simply the *association* of the virial path with the bond path at the level of the quantum state. It indicates a congruence between the topology and topography of the electron density and of the virial field. Nowhere does this reverse the logical order.

The local virial theorem connects the Laplacian of the density to kinetic and potential energy densities [6]:

$$-\frac{\hbar^2}{4m}\nabla^2 \rho(\mathbf{r}) = 2G(\mathbf{r}) + V(\mathbf{r}), \tag{8}$$

while the total electronic energy density is given by

$$H(\mathbf{r}) = G(\mathbf{r}) + V(\mathbf{r}). \tag{9}$$

These quantities help to classify bonding when evaluated at a bond critical point (bcp), as is well-known [6]. Shared-shell interactions typically show high $\rho_{bcp}$, negative $\nabla^2\rho_{bcp}$, and negative $H_{bcp}$. Closed-shell interactions typically show lower $\rho_{bcp}$, positive $\nabla^2\rho_{bcp}$, and small positive or slightly negative $H_{bcp}$ depending on the system. These signs and magnitudes are descriptors ($D$'s) of the density and energy-density fields and not additional causes external to the state.

The atomic virial theorem in QTAIM leads to a total atomic energy that is simply the negative of its kinetic energy, i.e.:

$$E(\Omega) = -T(\Omega) \tag{10}$$

for an atom in a molecule at equilibrium, when the virial relation is satisfied [6]. In this case, the molecular energy is expressed as the sum of unique well-defined atomic (virial) energy as [6]:

$$E_{\text{total}} = \sum_{\Omega} E(\Omega). \tag{11}$$

This last expression allows one to identify *which atoms are relatively stabilized or destabilized between two structures*. It is a powerful analytical tool but, and importantly, without making either "the bond" nor "the bond path" separate causal agents. Eqn (10) can be used to pinpoint *where* stabilization is localized within the quantum state at an atomic level of description. This is essential to understand the concept of the hydrogen–hydrogen bonding which is still debated from time to time.

**Hydrogen–hydrogen bonding: an often-misconstrued test case**

Hydrogen–hydrogen bonding [18–20] is an example that challenges simple intuition. Close H···H contacts are often described as "steric repulsions". Yet there is neither a $\hat{B}_{\text{steric repulsion}}$ operator nor a steric term in the Hamiltonian. Meanwhile, QTAIM shows that such contacts are *locally stabilizing* and *associated with bond paths* [18–20].

Hydrogen–hydrogen bonding differs from dihydrogen bonding. Dihydrogen bonding involves a protonic hydrogen and a hydridic hydrogen [21]. Hydrogen–hydrogen bonding involves two neutral, nearly neutral, or similarly charged hydrogen atoms (see discussion in Ref. [21]). H–H bond paths occur between ortho hydrogens in systems such as planar biphenyl, angular polybenzenoids, and crowded hydrocarbons. This interaction is almost universally associated with the relative lowering of the virial energies of the bonded hydrogen atoms [18–20].

Planar biphenyl is a transition state relative to the twisted equilibrium geometry. Classical language often says that ortho H···H "repulsion" destabilizes the planar form and "cause" the structure to twist. Repulsion implies a *force* acting outwards along the internuclear axis. Such force cannot exist by construction since the geometry has been optimized (Fig. 1). Virial energies reveal that each of the four ortho hydrogens is locally stabilized in the planar structure by about 8 kcal.mol$^{-1}$ each while the two junction carbons are destabilized by *ca*. 22 kcal mol$^{-1}$ each [19]. The net barrier is equal to the sum of the changes in the virial energies of all the atoms on biphenyl with these two groups being the dominant ones. As a result, the H–H contacts are *locally stabilizing* despite that the whole planar structure is *globally destabilized*.

It is important to emphasize the distinction between *local* relative stabilization from *global* stability, and *descriptor(s)* from *cause(s)*. The H–H bond path *accompanies* a locally stabilizing pattern, but the global energy difference is determined by the total state. *The bond path does not act as an independent causal entity, it only reveals the location of relative stabilization* [22].

**Bonding, Causation, and the Hierarchy of Chemical Description**

Scerri does not support the straightforward reduction of chemistry to quantum mechanics [23–25], despite the well-known 1929 assertion by Dirac [26] that

> "*The underlying physical laws necessary for the mathematical theory of a large part of physics and* the whole of chemistry [my emphasis] *are thus completely known, and the difficulty is only that the exact application of these laws leads to equations much too complicated to be soluble.*"

Contrary to Dirac's view, chemistry employs its own concepts, classifications, and explanatory language, which, while grounded in quantum mechanics, are not reducible to it in any simple sense. Hoffmann has stressed this position a number of times [27,28]. This is also aligned with the ontological autonomy of the chemical domain expressed by Lombardi and Labarca [29].

Michael Polanyi also opposed the reduction of a complex (chemical) system in favor of a hierarchy of independent descriptions. Higher-level organization imposes initial and boundary conditions on lower-level physical processes ("dual control") such that form-and-function, while not violating the underlying laws of physics, are not determined by them [30,31]. Although a molecule is completely specified by its quantum state governed by the Hamiltonian, chemists interpret this state through higher-level constructs such as bonding patterns, functional groups, and reaction mechanisms. These constructs do possess genuine predictive power. Bonding is inferred from the integration of multiple empirical descriptors such as internuclear distances, bond paths, interatomic zero-flux surfaces, energetic criteria, spectroscopic signatures, and reactivity patterns. This is precisely what Polanyi introduces as the tacit integration [32].

Terms such as "bond", "stabilize", and "cause" acquire their meaning by their use in practice [33,34]. These terms are descriptive and predictive but when used beyond this role, *e.g.*, when "the bond" is reified as a causal agent acting on the system, this is not simply imprecise wording but a category error where a derived state-dependent descriptor is misidentified as a primitive cause. The issue is, therefore, not the use of bonding language, which is indispensable, but rather its uncritical elevation from a descriptive shorthand to an ontological or causal claim [22].

In the first page of a well-known 1912 paper, Bertrand Russell warns against naïve causal language in fundamental science. He makes the following provocative statement [35]

"*The law of causality, I believe, like much that passes muster among philosophers, is a relic of a bygone age, surviving, like the monarchy, only because it is erroneously supposed to do no harm.*"

Quantum mechanics has a probabilistic character and, hence, a less clear cause-and-effect structure. Again, a cause must be independently manipulable; chemical bonds fail this criterion since any "change in a bond" necessarily entails a change in the underlying quantum state [14]. Taken together all this means that causal language involving bonding is, at best, model-relative and derivative [36–38].

Chemists have a sophisticated predictive machinery used daily in thousands of laboratories around the world to synthesize new materials and drugs and to predict reactivity. But prediction *is not* explaining, as the title of René Thom's book suggest [39]. Thom, the inventor of the *Catastrophe Theory* [40-42] which underpins the topological ingredients of QTAIM, defines prediction as *what works operationally*; and adds that this predictive power, in itself, is not a license for a causal explanation. On page 34 [39], he replies to his interviewer, "*c'est autour de cela que tourne le problème des modèles catastrophistes*" ["*It is around this that the problem of catastrophe models revolves*"]. In other words, prediction and explanation are different and Catastrophe Theory is one that affords the latter rather than the former.

Classical bonding models (steric repulsion, hydrogen bond stabilization of a conformer, VSEPR, EDA, NCI, etc.) are extraordinarily successful as predictors and classifiers of stable structures, yet their success should not be misconstrued as conferring a causal or explanatory status. The "reasons" a predictive model, such Gillespie and Nyholm's VSEPR model [43,44], works emerges only when extracts from quantum mechanics the Laplacian of the electron density [45]. Only then does the VSEPR transitions into both a predictive *and* an explanatory model, or rather "a theory".

**Hydrogen bonds do not cause protein stability (in condensed phases)**

The phrase "proteins are stabilized by hydrogen bonds" is both useful and misleading. It is useful because hydrogen-bonding patterns *correlate strongly* with secondary structure, folding, recognition, and active-site organization. It is misleading because the free energy of folding includes many terms: intramolecular electrostatics, exchange, dispersion, polarization, charge transfer, solvation, conformational entropy, hydrophobic reorganization, ionization, protonation states, and environmental effects. A hydrogen bond is a pattern within this whole balance. It is not an independent stabilizing agent. The quantum and statistical-mechanical state of the protein coupled with its solvent environment has a free-energy landscape with basins. Within those basins, certain donor–acceptor geometries, density patterns, and electrostatic arrangements are classified as hydrogen bonds.

In a solvated state, the arguments presented above must be restated statistically rather than structurally. A folded conformation is not reached by the interplay of "stabilizing hydrogen bonds" and "destabilizing steric repulsions", but by the free-energy balance over an ensemble of microstates,

$$G = -k_B T \ln Z, \tag{12}$$

where the canonical partition function $Z$ contains energetic and entropic contributions from intramolecular interactions, solvent, ions, conformational freedom, and hydrophobic effects. Hydrogen bonds, steric contacts, salt bridges, and hydrophobic contacts are therefore best treated as descriptors of recurring structural motifs within a free-energy basin, not as primitive causes of that basin.

The statistical-mechanical description of protein folding reflects a balance of enthalpic and entropic terms rather than a single structural feature [46–48]. The statement "hydrogen bonds stabilize a protein" should be read as model-relative shorthand for a contribution to an ensemble free energy, not as a fundamental causal claim.

The Gibbs energy $G$ is ultimately determined by the Hamiltonian of the complete system *via* the partition function and its value depends on the full ensemble of microstates rather than on any structural descriptor(s). Bonding descriptors remain, therefore, state-dependent summaries of that ensemble and not primitive causal terms in the Hamiltonian.

**Conclusion**

Chemical bonding is indispensable in the chemical discourse, but its logical status must be stated precisely. The molecular Hamiltonian contains no bonding terms, and quantum mechanics defines no "bonding operator". The Born–Oppenheimer approximation allows us to construct potential energy hypersurfaces and locate stationary structures. Once a stationary structure has been found then one can calculate descriptors such as bond paths, virial paths, ELF basins, NCI surfaces, EDA components, *etc*. All these are "readings" of the quantum state but none constitutes a primitive causal entity.

Bonding models classify, correlate, and predict. Their success makes them indispensable, but predictive utility does not confer causal or fundamental explanatory status. The persistent attribution of stability to "bonds" and instability to "steric repulsion" reflects either circular reasoning and asymmetrical causal attribution, respectively. More precisely, it is a category error in which a derived state-dependent descriptor is misidentified as a primitive cause.

The linguistic point is not trivial. As emphasized by Bader, "bonding" is more appropriate than "bond" [22]. The noun invites reification; the verb preserves interaction. This distinction mirrors the conceptual one: bonding is a pattern within the quantum state, not an object acting upon it. Accordingly, the correct formulation is simple and rigorous: *Bonding* accompanies *quantum states and organizes chemical reasoning but does not stand outside the state as the primitive source of its stability*.

**Acknowledgements**

The author gratefully acknowledges the lasting influence of his mentor and friend, Professor Richard F. W. Bader (1931–2012), to whose memory this Essay is dedicated. Financial support from the Natural Sciences and Engineering Research Council of Canada (NSERC), the Canada Foundation for Innovation (CFI), and Mount Saint Vincent University is also gratefully acknowledged.